\documentclass[                  
 	,draft
               ]{aipproc}
               
\layoutstyle{6x9}

\begin{document}
%%%%%%%%%%%%%%%%%%%%%%%%%%%%%%%%

\title{Conformal diagrams for the gravitational collapse of a spherically symmetric dust cloud}

% 04.20.-q: Classical GR
% 04.25.-g: Approximation methods; equations of motion
% 04.25.Dm: Numerical relativity
% 04.40.-b: Self-gravitating systems, continuous media and 
%           classical fields in curved spacetime
% 04.70.-s: Physics of black holes
% 97.60.Lf: Astronomy: Late stage of star evolution: black holes
% Complete list see \texttt{http://www.aip..org/pacs/index.html}
           

\author{N\'estor Ortiz}{
address={Instituto de F\'{\i}sica y Matem\'aticas,
Universidad Michoacana de San Nicol\'as de Hidalgo\\
Edificio C-3, Ciudad Universitaria, 58040 Morelia, Michoac\'an, M\'exico.}
} 

\author{Olivier Sarbach}{
address={Instituto de F\'{\i}sica y Matem\'aticas,
Universidad Michoacana de San Nicol\'as de Hidalgo\\
Edificio C-3, Ciudad Universitaria, 58040 Morelia, Michoac\'an, M\'exico.}
}

%%%%%%%%%%%%%%%%%%%%%%%%%%%%%%%%
\begin{abstract}
%%%%%%%%%%%%%%%%%%%%%%%%%%%%%%%%
We present an algorithm for the construction of conformal coordinates in the interior of a spherically symmetric, collapsing matter cloud in general relativity. This algorithm is based on the numerical integration of radial null geodesics. As an application we generate conformal diagrams for collapsing spherical dust clouds and analyze the causal structure of the resulting spacetimes.
\end{abstract}

\maketitle

%%%%%%%%%%%%%%%%%%%%%%%%%%%%%%%%%%%%%%%%%%%%%%
\section{Introduction}
\label{Sect:Intro}
%%%%%%%%%%%%%%%%%%%%%%%%%%%%%%%%%%%%%%%%%%%%%%

General relativity predicts that sufficiently massive, cold fluid stars cannot exist in hydrostatic equilibrium and, hence, such stars must undergo complete gravitational collapse (see \cite{HawkingEllis-Book,Wald-Book} and references therein). If a trapped surface forms during the collapse, then the singularity theorems \cite{HawkingEllis-Book} also imply that the final state includes a spacetime singularity. One of the interesting questions is whether this singularity is censured, that is, if it is hidden inside a black hole, or if it is naked in the sense that light rays emanating from the singularity can reach observers which are arbitrary far from the collapsing star. According to the weak cosmic censorship conjecture \cite{rP69} the former case must occur. A stronger form of this conjecture \cite{rP79} states that the future Cauchy development of generic, regular, asymptotically flat initial data gives rise to a globally hyperbolic spacetime, implying that the singularities in this spacetime are not visible to any observer.

On the other hand, it is known that the collapse of a spherical dust cloud leads to the formation of shell-focusing singularities which may be visible to local observers, and even to distant observers \cite{dElS79,dC84,rN86}. However, it is not clear whether or not these examples provide a serious threat to the cosmic censorship conjecture. Indeed, it remains to be seen if the nature of the singularity is unchanged under small perturbations of the initial data and the equation of state. For reviews and a recent book on this subject we refer the reader to \cite{rW97,pJ00} and \cite{Joshi-Book}, respectively.

In this article, we describe a numerical method for the construction of conformal diagrams in the spherical symmetric dust collapse. We apply this method to different initial data and analyze whether the resulting singularity is censored or naked. Our method should be generalizable to the spherical collapse of more general matter and is expected to be useful as a tool for an understanding of the dependency of the nature of the singularity (censored or naked) on the initial data.

%%%%%%%%%%%%%%%%%%%%%%%%%%%%%%%%%%%%%%%%%%%%%%
\section{The model}
\label{Sect:Model}
%%%%%%%%%%%%%%%%%%%%%%%%%%%%%%%%%%%%%%%%%%%%%%

As a simple collapse model we consider a self-gravitating, spherically symmetric dust cloud. In terms of co-moving, synchronous coordinates \cite{MTW-Book}
$(\tau,R)$, this model is described by the one-dimensional mechanical system
\begin{equation}
\frac{1}{2} \dot{r}(\tau,R)^2 + V(r(\tau,R), R) = E(R),\qquad
V(r,R) := -\frac{m(R)}{r},
\label{Eq:1DMechanical}
\end{equation}
where $R$ labels the different dust shells and where the function $\tau\mapsto r(\tau,R)$ describes the evolution of the areal radius along dust shell $R$ as a function of proper time. It is assumed that $r(0,R) = R$ initally; that is, $R$ coincides with the areal radius of the shell $R$ at the initial time $\tau=0$. The energy function $E(R) = v_0(R)^2/2 - m(R)/R$ is determined by the initial velocity and density profiles $v_0(R) := \dot{r}(0,R)$ and $\rho_0(R)$, respectively, where $m(R)$ is the Misner-Sharp mass function \cite{cMdS64} obtained from
\begin{equation}
m(R) = 4\pi G\int\limits_0^R \rho_0(s) s^2 ds,
\label{Eq:DustMass}
\end{equation}
with Newton's constant $G$. Once the function $r(\tau,R)$ has been determined, the time-evolution of the metric ${\bf g}$, the four-velocity ${\bf u}$ and the density $\rho$ are obtained according to the following formulae:
\begin{eqnarray}
{\bf g} &=& -d\tau^2 + \frac{r'(\tau,R)^2}{1 + 2 E(R)}\; dR^2
 + r(\tau,R)^2(d\vartheta^2 + \sin^2\vartheta\, d\varphi^2),
\label{Eq:MetricSol}\\
{\bf u} &=& \frac{\partial}{\partial\tau}\; , \qquad
\rho(\tau,R) = \rho_0(R)\left( \frac{R}{r(\tau,R)} \right)^2\frac{1}{r'(\tau,R)}\; .
\label{Eq:FluidSol}
\end{eqnarray}
Here and in the following, a dot and a prime denote partial differentiation with respect to $\tau$ and $R$, respectively.

Eq. (\ref{Eq:1DMechanical}) can be solved exactly \cite{rN86}. Since we are interested in the collapse of a dust cloud with positive density whose initial data does not contain any apparent horizons, we make the following assumptions
\begin{equation}
\rho \geq 0,\qquad
v_0\leq 0,\qquad
-\frac{1}{2} < E < 0.
\label{Eq:Assumptions1}
\end{equation}
In this case, the solution to (\ref{Eq:1DMechanical}) is given by
\begin{displaymath}
r(\tau,R) = \frac{R}{p(R)} F^{-1}\left( F(p(R)) + \sqrt{c(R)p(R)^3}\tau \right),
\end{displaymath}
where $p(R) := E(R)/V(R,R)\in (0,1]$ is the ratio between the total and initial potential energies, $c(R) := 2m(R)/R^3$ is proportional to the mean density within the dust shell $R$ and where the strictly decreasing function $F$ is given by $F: [0,1] \to [0,\pi/2], x\mapsto \sqrt{x(1-x)} + \arccos\sqrt{x}$. In the following, we also assume that the functions $p$ and $c$ are subject to the inequalities
\begin{equation}
p' \geq 0, \qquad c'\leq 0.
\label{Eq:Assumptions2}
\end{equation}
These conditions are fulfilled, for instance, if the density profile is decreasing and the initial velocity is zero. They imply the absence of {\em shell-crossing} singularities \cite{rN86}. There remains a {\em shell-focusing}\footnote{See \cite{pSaL99} for the distinct physical properties of shell-crossing and shell-focusing singularities.} singularity at points where $r(\tau,R)/R = 0$, in which case
\begin{displaymath}
\tau = \tau_s(R) = \frac{\frac{\pi}{2} - F(p(R))}{\sqrt{c(R)p(R)^3}}.
\end{displaymath}
This corresponds to a curvature singularity since the Ricci scalar diverges as a consequence of Einstein's field equations and the divergence of $\rho$ at those points.

In the next section we describe a method for generating conformal diagrams for the spacetime metric (\ref{Eq:MetricSol}) subject to the assumptions (\ref{Eq:Assumptions1},\ref{Eq:Assumptions2}). By definition, in- and outgoing radial null geodesics in such a diagram are given by straight lines with slope $-1$ and $1$, respectively, which nicely displays the causal structure of the underlying spacetime. In particular, we are interested in determining whether the shell-focusing singularity is censored or naked.

%%%%%%%%%%%%%%%%%%%%%%%%%%%%%%%%%%%%%%%%%%%%%%
\section{Conformal diagrams}
\label{Sect:Diagrams}
%%%%%%%%%%%%%%%%%%%%%%%%%%%%%%%%%%%%%%%%%%%%%%

In this section we describe a method for constructing conformal coordinates $(T,X)$ for the spacetime described by Eq. (\ref{Eq:MetricSol}). We assume that the cloud has a finite initial radius $R_1 > 0$. For $R > R_1$ the density is zero and according to Birkhoff's theorem the spacetime metric (\ref{Eq:MetricSol}) must reduce to the Schwarzschild solution whose causal structure is known. Therefore, it is sufficient to construct conformal coordinates for the region $\Omega := \{ (\tau,R) : 0 \leq R \leq R_1, 0\leq \tau < \tau_s(R) \}$ inside the dust cloud. Such coordinates can be obtained by first solving the advection equations
\begin{equation}
\dot{U} = -\gamma U',\qquad
\dot{V} = \gamma V',\qquad
\gamma := \frac{\sqrt{1 + 2E}}{r'},
\label{Eq:Advection}
\end{equation}
with appropriate initial and boundary data for the null coordinates $U$ and $V$ and then setting $T = (V + U)/2$, $X = (V - U)/2$. Eqs. (\ref{Eq:Advection}) in turn can be reduced to ordinary differential equations by the method of characteristics: the functions $V$ and $U$ are constant along the integral curves belonging to the null vector fields
\begin{displaymath}
k^{(in)} :=  \frac{\partial}{\partial\tau} - \gamma\frac{\partial}{\partial R},\qquad
k^{(out)} :=  \frac{\partial}{\partial\tau} + \gamma\frac{\partial}{\partial R},
\end{displaymath}
respectively. These integral curves describe in- and outgoing radial null geodesics. Since $\gamma > 0$ on $\Omega$, the integral curves belonging to $k^{(in)}$ cannot cross at points away from the singularity. The same statement holds for the integral curves to $k^{(out)}$. Furthermore, a careful argument \cite{dC84,rN86} shows that under appropriate regularity assumptions on the initial data $\rho_0$ and $v_0$, the ingoing radial null geodesics cannot cross on the singularity ${\cal S}:= \{ (\tau,R) : 0 \leq R \leq R_1 : \tau =\tau_s(R) \}$, and that the outgoing radial null geodesics cannot cross on ${\cal S}$ except at the point $\tau_s^0 = (0,\tau_s(0))$. Therefore, we can construct $U$ and $V$ by specifying their values on the singularity ${\cal S}$, by specifying $V$ at the center line ${\cal C} := \{ (\tau,R) : R = 0, 0\leq \tau < \tau_s^0 \}$ and by specifying $U$ at the surface of the cloud, ${\cal T} := \{ (\tau,R) : R = R_1, 0\leq \tau < \tau_s(R_1) \}$.

We first apply our method to the case of a homogeneous dust cloud with zero initial velocity and then describe a numerical algorithm which is applied to the generic case.

\subsection{Special case of a homogeneous cloud with zero initial velocity}

\begin{figure}
\includegraphics[height=.45\textheight]{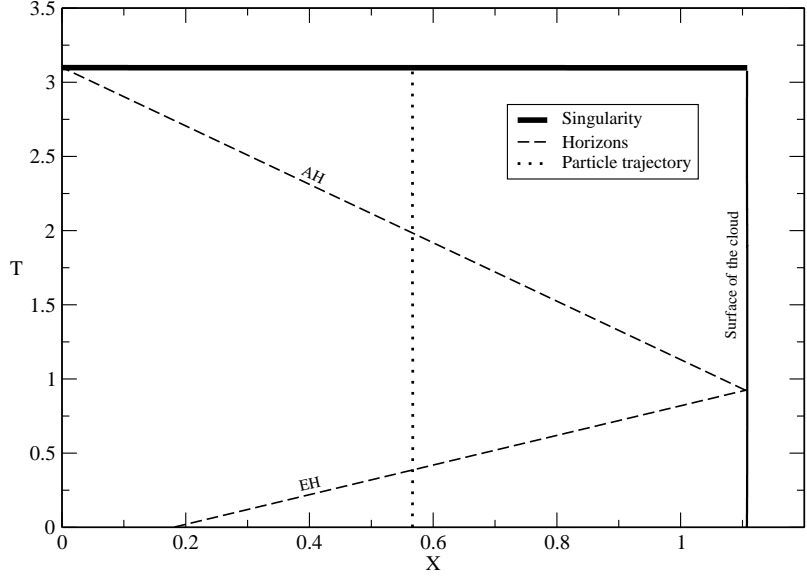}
\caption{Conformal diagram for a homogenous dust cloud with zero initial velocity, $c_0 = 0.75$ and $R_1=1$. The lines denoted by "AH" and "EH" refer to the apparent and event horizons, respectively.}
\label{Fig:ConfHomog}
\end{figure}

For a dust cloud with homogeneous density and zero initial velocity, $c(R) = c_0$ is constant and $p=1$ on $\Omega$. The assumptions (\ref{Eq:Assumptions1}) require $0 \leq c_0 < 1$. In this case, $\gamma = \sqrt{1 - c_0 R^2}/F^{-1}(\sqrt{c_0}\tau)$ and the coordinate transformation
\begin{displaymath}
\sqrt{c_0}\tau = F\left( \cos^2\left(\frac{T}{2}\right) \right), \qquad
\sqrt{c_0} R = \sin(X),\qquad
0 \leq T < \pi, 0\leq X \leq \arcsin(\sqrt{c_0} R_1)
\end{displaymath}
gives
\begin{displaymath}
k^{(in)} = \frac{\sqrt{c_0}}{\cos^2\left( \frac{T}{2} \right)}
\left( \frac{\partial}{\partial T} - \frac{\partial}{\partial X} \right),\qquad
k^{(out)} = \frac{\sqrt{c_0}}{\cos^2\left( \frac{T}{2} \right)}
\left( \frac{\partial}{\partial T} + \frac{\partial}{\partial X} \right).
\end{displaymath}
Therefore, a solution to the advection equations (\ref{Eq:Advection}) is $U = T+X$ and $V=T-X$. The corresponding conformal diagram for the choice $c_0 = 0.75$ is displayed in Figure~\ref{Fig:ConfHomog}. Here, the apparent horizon is the boundary between the points for which the areal radius $r$ increases along outgoing radial null geodesics, $k^{(out)}[r] > 0$, and the points for which $k^{(out)}[r] < 0$. A small calculation reveals that this surface is determined by the simple equation $r(\tau,R) = 2m(R)$. For the homogenous case with zero initial velocity considered here this reduces to $T = \pi - 2X$.
Since outside the cloud the apparent horizon coincides with the event horizon, the event horizon inside the cloud is determined by the outgoing null ray passing through the intersection of the apparent horizon with the surface of the cloud.

\subsection{The generic case}

For the generic case, we apply the algorithm described above in order to determine the functions $U$ and $V$ on the inside region $\Omega$. To this purpose, we select a point $p$ inside $\Omega$. Then, we numerically integrate the future-directed in- and outgoing radial null rays emanating from $p$. The ingoing ray ends at either the center of the cloud ${\cal C}$ or at the singularity ${\cal S}$. The outgoing ray ends at ${\cal S}$ or at the surface of the cloud, ${\cal T}$. From these two end points we define the coordinates $(U,V)$ associated to the point $p$ according to
\begin{displaymath}
\left. V \right|_{\cal C} = \frac{\tau}{\tau_s(0)},\qquad
\left. U \right|_{\cal T} = \frac{\tau}{\tau_s(R_1)} - 1,\qquad
\left. V \right|_{\cal S} = 1 + \frac{R}{R_1},\qquad
\left. U \right|_{\cal S} = 1 - \frac{R}{R_1}.
\end{displaymath}
Notice that this choice implies that the singularity ${\cal S}\setminus\tau_s^0$ is determined by the line $T=1$ and $0 < X=R/R_1 \leq 1$ in the conformal diagram. On the other hand, it turns out there are infinitely many outgoing null rays emanating from $\tau_s^0$. These null rays have the same value of $V=1$ and different values for $U$. Therefore, the point $\tau_s^0$ unfolds to a portion of the straight line $T=1-X$ in the conformal diagram. Two examples are shown in Figures~\ref{Fig:ConfOculta}~and~{\ref{Fig:ConfDesnuda}.
\begin{figure}
\includegraphics[height=.45\textheight]{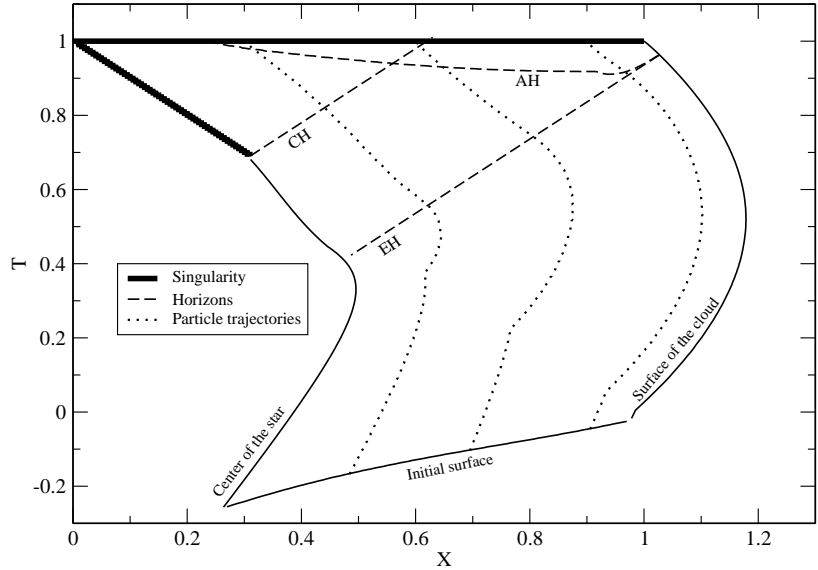}
\caption{Conformal diagram for the model described in Eq. (\ref{Eq:cpChoice}) with the parameter choice $c_0 = 0.65$, $a=0.7$, $p_0=0.1$ and $p_1=0.9$. The lines denoted by "AH", "EH" and "CH" refer to the apparent, event and Cauchy horizons, respectively. In this case, the singularity $\tau_s^0$ is hidden inside the black hole region since all light rays emanating from it end at the singularity $T=1$.}
\label{Fig:ConfOculta}
\end{figure}
\begin{figure}
\includegraphics[height=.45\textheight]{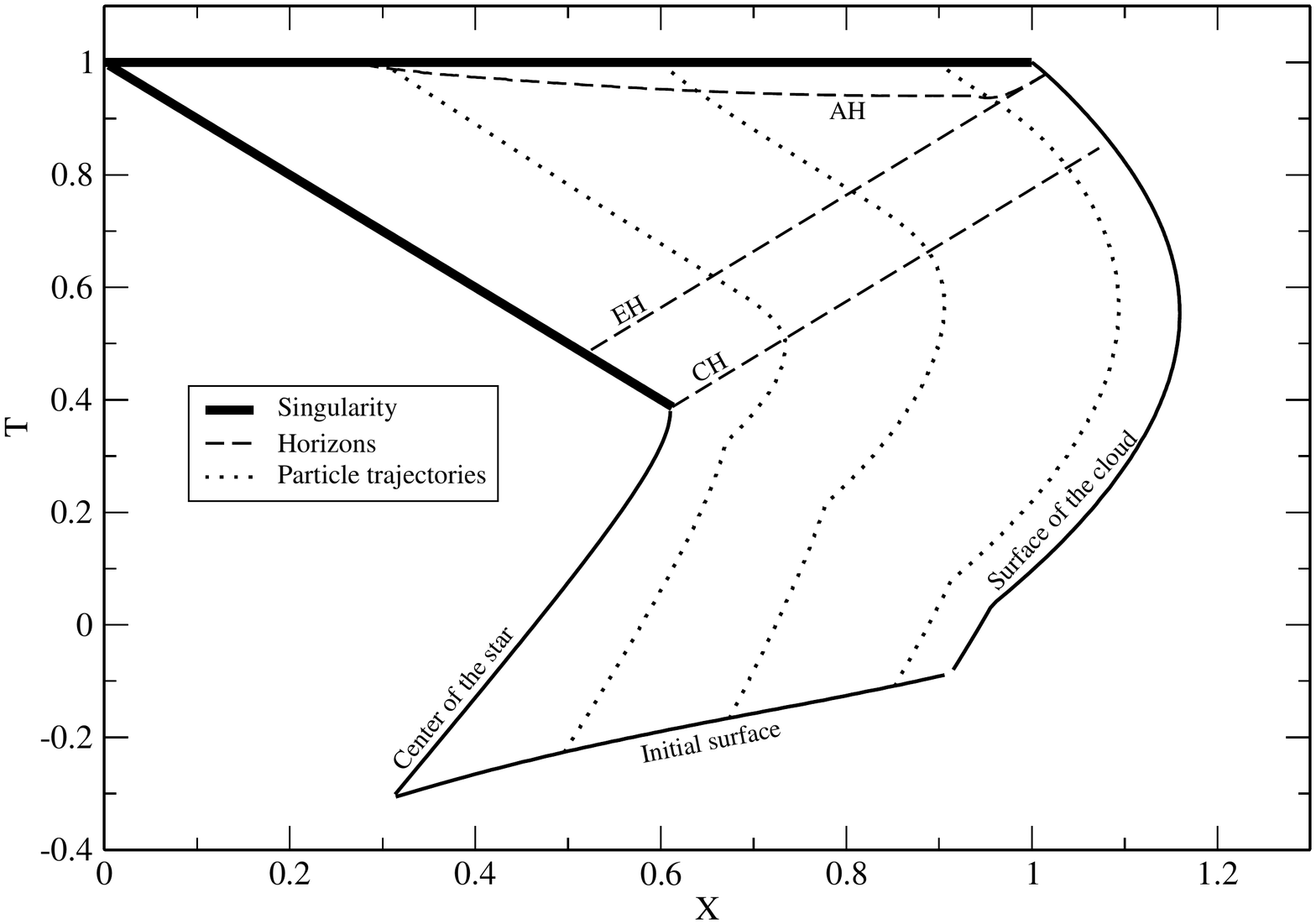}
\caption{Conformal diagram for the model described in Eq. (\ref{Eq:cpChoice}) with the same parameter choice as in the previous figure except that $c_0 = 0.45$. In this case, there exists light rays emanating from $\tau_s^0$ which arrive at the surface of the cloud {\em earlier} than the apparent horizon. Therefore, a portion of this singularity is visible to 
distant observers {\em outside} the black hole region.}
\label{Fig:ConfDesnuda}
\end{figure}
In the first case, the singularity is hidden inside the black hole while in the second case a portion of $\tau_s^0$ is visible from future null infinity. The results are based on the following choice for the functions $c$ and $p$:
\begin{equation}
c(R) = c_0\left[ 1 - \frac{6}{5} a\left( \frac{R}{R_1}\right)^2 
+ \frac{3}{7}(2a-1)\left(\frac{R}{R_1}\right)^4 \right],\qquad
p(R) = p_0 + (p_1 - p_0)\frac{R}{R_1},
\label{Eq:cpChoice}
\end{equation}
where the parameters $c_0$, $a$, $p_0$ and $p_1$ are subject to the inequalities 
\begin{displaymath}
0 \leq a \leq 1,\quad
0 < c_0 < \frac{7}{6a+4},\qquad
0 < p_0 \leq p_1 \leq 1,
\end{displaymath}
which guarantee the satisfaction of the assumptions (\ref{Eq:Assumptions1},\ref{Eq:Assumptions2}). Furthermore, $c(R)$ is such that the density $\rho$ converges to zero as $R\to R_1$ approaches the surface of the cloud.

For the numerical integration of the light rays we first rescale the vector fields $k^{(in)} $ and $k^{(out)}$ by the factor $(1 + \gamma^2)^{-1/2}$ and then solve the system of ordinary differential equation
\begin{equation}
\frac{d\tau}{d\lambda} = \frac{1}{\sqrt{1 + \gamma(\tau,R)^2}}, \qquad
\frac{dR}{d\lambda} =  \frac{\gamma(\tau,R)}{\sqrt{1 + \gamma(\tau,R)^2}}.
\label{Eq:ODE}
\end{equation}
This rescaling corresponds to a reparametrization of the light curves and offers the advantage of avoiding stiff coefficients. Therefore, adaptive or implicit time integration is not necessary. We numerically integrate Eq. (\ref{Eq:ODE}) using a fourth-order Runge-Kutta time integrator (see, for instance, \cite{Recipies-Book}) with fixed step size $h$. The 
results for the apparent horizon corresponding to different values for $h$ are shown in Figure~\ref{Fig:AHConv}, and indicate convergence of our method. We have also verified the convergence of the earliest point of the image of $\tau_s^0$ in the conformal diagram, from which the Cauchy horizon emanates.
\begin{figure}
\includegraphics[height=.45\textheight]{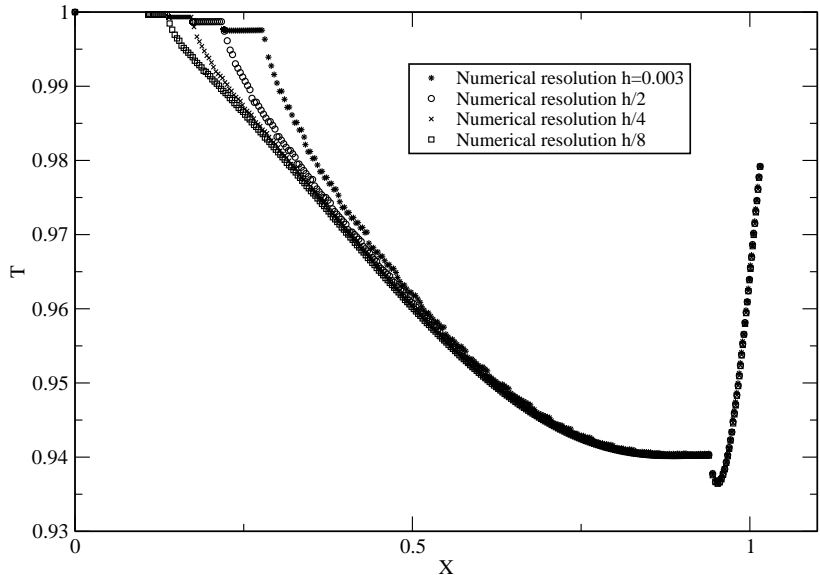}
\caption{The location of the apparent horizon in the conformal diagram for different resolutions. The corresponding convergence factor oscillates about a value close to three, indicating at least first order convergence. The initial density and velocity profiles are the same as in the previous figure.}
\label{Fig:AHConv}
\end{figure}

%%%%%%%%%%%%%%%%%%%%%%%%%%%%%%%%%%%%%%%%%%%%%%
\section{Conclusions}
\label{Sect:Conclusions}
%%%%%%%%%%%%%%%%%%%%%%%%%%%%%%%%%%%%%%%%%%%%%%

We presented a simple numerical method for determining the conformal diagram inside a collapsing, spherical dust cloud with generic initial density and velocity profiles. This provides a valuable tool for studying under which conditions the singularity is hidden inside a black hole and under which conditions it is visible from future null ifninity. Our method should also be applicable to spacetimes describing the spherically symmetric collapse of clouds with nonzero pressure since it is based on the integration of radial null rays. However, unlike to present dust case, the spacetime metric for generic initial data can hardly be obtained in analytic form for realistic equations of state, and a numerical integration is probably necessary for obtaining the solution to the field equations.

%%%%%%%%%%%%%%%%%%%%%%%%%%%%%%%%%%%%%%%%%%%%%%
\begin{theacknowledgments}
%%%%%%%%%%%%%%%%%%%%%%%%%%%%%%%%%%%%%%%%%%%%%%
It is a pleasure to thank Jos\'e Gonz\'alez, Francisco Guzm\'an and Thomas Zannias for interesting discussions. This work was supported in part by Grants No. CIC 4.19 to Universidad Michoacana and CONACyT 61173.
\end{theacknowledgments}

\bibliographystyle{aipproc}  
\bibliography{refs_collapse}
%%%%%%%%%%%%%%%%%%%%%%%%%%%%%%%%%%%%%%%%%%%%%%

\end{document}